\documentstyle[twoside,fleqn,espcrc2]{article}

\newcommand{\Det}{{\rm Det}}
\newcommand{\Tr}{{\rm Tr}}
\newcommand{\MdM}{M^{\dagger}\!M}
\newcommand{\NdN}{N^{\dagger}\!N}
\newcommand{\Plaq}{\left< P \right>}

\begin{document}

\title{The Numerical Estimation of the Error Induced
        by the Valence Approximation\thanks{Talk presented
       by James Sexton}}

\author{James C. Sexton%
       \address{School of Mathematics, Trinity College,\\
               Dublin 2, Ireland}%
       \thanks{Travel supported in part by EC-HCM Contract CHRX-CT92-0051.}
       and
       Donald H.\ Weingarten%
       \address{IBM, T. J. Watson Research Center,\\
               P. O. Box 218, Yorktown Heights,\\
               NY 10598, U.S.A}}

\begin{abstract}
We describe a systematic expansion for full QCD.  The leading
term in the expansion gives the valence approximation.  The
expansion reproduces full QCD if an infinite number of
higher terms are included.
\end{abstract}

\maketitle\section{INTRODUCTION}

The observation that recent valence approximation predictions are quite
close to experiment strongly suggests that, at least for some
observables, the systematic error arising from this approximation is
quite small.  It then seems natural to ask whether an independent
quantitative estimate can be made of the systematic error of the valence
approximation.  In this article, we describe an expansion for full QCD
\cite{Sexton94} which has the valence approximation as its leading term,
reproduces full QCD exactly if an infinite number of higher terms
are included, allows the numerical calculation of lattice QCD
quantities for any finite truncation of the expansion, and allows
the estimation of the error induced by any such finite truncation.

The basic idea behind the scheme we suggest is to develop an effective
action to approximate the log of the determinant of the fermion hopping
matrix.  The valence approximation replaces this term by a quantity
proportional to the pure gauge action.  Improved approximations can be
generated by replacing the log of the determinant by a more complicated
function of the links.  A useful set of such functions is generated by
sums of traces of products of links about closed paths on
the lattice, and our approximation scheme involves expanding the
determinant of the hopping matrix as a linear combination of these
functions.

\section{DEFINITIONS}

We will work with Wilson's formulation of lattice QCD.  If $M(\kappa)$
is Wilson's hopping matrix for a single flavor fermion with hopping
parameter $\kappa$, then our aim is to find an approximation for $\Det
M$ ($= \sqrt{\Det\MdM}$).  Numerical simulations of QCD work on finite
lattices, with $\kappa$ not too big.  For almost all configurations on
such lattices $\MdM$ is positive definite and has finite, strictly
positive minimum $\lambda_{\rm min}$ and maximum $\lambda_{\rm max}$
eigenvalues. Also, it is simple to see (for example by considering a
hopping parameter expansion) that $\Det\left(\MdM\right)$ can be
expressed as a finite linear combination of sums of traces of products
of links about closed paths.  These two observations allow us to define
$\log\left(\Det\left(\MdM\right)\right)$ as an infinite series,
convergent with respect to a suitably defined norm \cite{Sexton94}
\begin{equation}
\label{Eqn:TrLn}
\log\Det\left(\MdM\right) =
\Tr\log\MdM = \sum_{i=0}^{\infty} a_i S_i(U)
\end{equation}
where the $S_i(U)$ are a maximal linearly independent set of
functions which involve sums of traces of products of links
about closed paths.

Crucial to obtaining a useful truncation of the expansion in
Eq.~(\ref{Eqn:TrLn}) is the ordering assigned to the sequence of
$S_i(U)$.  We choose to partially order these functions by the length of
the closed paths involved.  For what follows we need explicit forms only
for the constant (path length 0) term
\begin{equation}
S_0(U) = 1
\end{equation}
and the path length 4 term
\begin{equation}
S_1(U)  =  \frac{1}{3} \sum_{\Box} {\rm Re} \Tr U_{\Box}
\end{equation}
involving a sum over all plaquettes of the real
part of the trace of the product $U_{\Box}$ of links
about a plaquette.

\section{APPROXIMATION SCHEME}

The series expansion for $\Tr\log\MdM$ defined in
(\ref{Eqn:TrLn}) allows us to develop a series of approximations
for full QCD expectation values of observables $F$.
If we include two equal mass flavors of fermions,
(the case most simply treated with exact algorithms),
we have
\begin{eqnarray}
\left< F \right> & = & Z^{-1} \int d\mu \; F \;
  \exp\left(\beta S_1\!+\!\Tr\log\MdM\right)
  \nonumber \\
& = & Z^{-1} \int d\mu \; F \;
  \exp\left(\beta S_1\!+\!\sum a_i S_i\right)
\end{eqnarray}
We approximate these expectation values by truncating
the series for $\Tr\log\MdM$.

If $L_n = \sum_{i=0}^{n} a_i S_i$ is such a truncated series, and if
$R_n = \Tr\log\MdM\! -\! L_n$ is the remainder after truncation, then we
define our approximate expectation $\left< F \right>_n$ after this
truncation by
\begin{equation}
\left< F \right>_n = Z^{-1} \int d\mu \; F \;
  \exp\left(\beta S_1\!+\! L_n\right)
\end{equation}
The systematic error induced by the truncation is
$\left<F\right> - \left<F\right>_n$, and if the correlations of $F$ and
$R_n$ are small, then an estimate of this error is
\begin{equation}
\label{Eqn:SystematicError}
\left<F\right> - \left<F\right>_n \approx
\left< \left( F - \left<F\right>_n \right)
       \left( R_n - \left<R_n\right>_n \right) \right>_n
\end{equation}

Still to be determined at this stage are values for the coefficients
$a_i$.  The norm implicit in Eq.~(\ref{Eqn:TrLn}) yields
\begin{equation}
\label{Eqn:FormulaForAs}
\lim_{n \rightarrow \infty}
\left< \left(\Tr\log\MdM - \sum_{i=0}^n a_i S_i\right)^2
\right> = 0.
\end{equation}
We can find approximate values for the coefficients
at any given truncation by minimizing
\begin{equation}
\label{Eqn:MinCondition}
\left< R_n^2 \right>_n =
\left< (\Tr\log\MdM - L_n)^2 \right>_n
\end{equation}
as a function of $a_0 \ldots a_n$.  In the limit $n\rightarrow\infty$
the coefficients $a_i$ will approach their exact values determined by
(\ref{Eqn:TrLn}) or equivalently (\ref{Eqn:FormulaForAs}).

\section{CALCULATING $\Tr\log\MdM$}

In order to evaluate the coefficients $a_i$ (by minimizing
(\ref{Eqn:MinCondition})), or to estimate the systematic error by
equation (\ref{Eqn:SystematicError}), we must be able to evaluate the
expectation value of $\Tr\log\MdM$ over a set of link configurations.
We use two different ideas to evaluate these expectations.  Firstly, we
estimate $\Tr\log\MdM$ as
\begin{equation}
\Tr\log\MdM = \frac{1}{N_\phi} \sum_{a=1}^{N_\phi}
\phi^{\dagger}_a \!\cdot\! \log\MdM \!\cdot\! \phi_a
\end{equation}
were $\phi_a$ are $N_\phi$ pseudofermion vectors each with independent
random Gaussian components.  Secondly, to calculate
$\log\MdM \cdot \phi$ we use a Chebyshev polynomial iterative scheme
\begin{eqnarray}
\log\MdM & =  &
  \sum_{i=0}^N b_i T^{*}_i\left( \frac{1-\MdM/\lambda_{\rm max}}
                                 {1-\lambda_{\rm min}/\lambda_{\rm max}}
                                 \right)
\nonumber \\
  & + & \delta \log\MdM
\end{eqnarray}
where the $T^{*}_i$ are Chebyshev polynomials,
the $b_i$ are constants which can be calculated using standard
methods for orthogonal polynomials \cite{Fox68}, and $\delta$ is bounded
by
\begin{equation}
\delta < 2 \exp\left( -2 N \sqrt{\lambda_{\rm min}/\lambda_{\rm max}} \right)
\end{equation}
Convergence of this series is determined by $\delta$, and since
$\log\MdM \!\cdot\! \phi$ must be calculated a number of times in order
to evaluate the trace we have found that the most efficient
implementation of these ideas is first to calculate $\lambda_{\rm min}$ and
$\lambda_{\rm max}$ for each configuration being analyzed
using a Lanczos algorithm \cite{Golub89}, then to evaluate
$\phi^{\dagger} \!\cdot\! \log\MdM \!\cdot\! \phi$ with a Chebyshev polynomial
approximation whose order $N$ is defined to keep
$ N \sqrt{ \lambda_{\rm min}/\lambda_{\rm max} } $ on each configuration
constant.  We have also found that it helps to precondition the
$\log\MdM$ calculation by premultiplying $M$ by the inverse of a
free hopping matrix $M_0(\kappa_0)$ which has hopping parameter
$\kappa_0$, and has all links set to the identity.
This preconditioner has no effect on the physics since,
if $N = M_0^{-1}(\kappa_0) M$, we have $\Det\NdN={\rm const}\times \Det\MdM$.
However, simple tuning of $\kappa_0$ reduced the work to calculate
the determinant in our example calculation by almost a factor of two.
Details of
our exact implementation of these ideas can be found in \cite{Sexton94}.

\section{EXAMPLE APPLICATION}

As a first demonstration of these ideas, we have studied a $6^4$
QCD simulation at $\beta = 5.7$ with two equal mass Wilson fermions
of hopping parameter $\kappa = 0.16$.  We considered
only the first two terms in our expansion
\begin{equation}
\Tr\log\NdN \rightarrow L_1 = a_0 + a_1 S_1(U).
\end{equation}
Observable expectations in this approximation have the
form
\begin{equation}
\left<F\right>_1 = Z^{-1} \int d\mu \; F \;
  \exp\left( (\beta\!+\! a_1) S_1(U)\!+\! a_0 \right)
\end{equation}
This expectation value is a valence approximation calculation
but we now obtain an explicit number for the shift in $\beta$
compensating for omission of the fermion determinant.
The coupling
at which we simulate becomes
$\beta_{\rm sim}\! =\!  \beta\! + \! \delta \beta$,
where $\delta\beta\! =\! a_1$ is the effective shift
in the coupling due to the fermion determinant.
To calculate $\delta\beta$ we generated 160 independent configurations
using the Cabbibo-Marinari-Okawa algorithm with $\beta_{\rm sim} = 5.7$.
On these configurations, we calculated $\Tr\log\NdN$
using 140 independent $\phi$'s per configuration.

The major result of all our analysis is a prediction for the constant
$\delta\beta$ which is generated by minimizing (\ref{Eqn:MinCondition}).
We found $\delta\beta = -0.261 \pm 0.014$.
We interpret this result to mean that a valence calculation at
$\beta = 5.7$ is approximately equivalent to a full QCD calculation at
$\beta=5.439$   ($5.700\! -\! 0.261$)
for two equal mass flavors of Wilson fermions with hopping parameter
$\kappa = 0.16$.  The agreement is approximate because
we have truncated the expansion of $\Tr\log\NdN$ at two
terms.

To confirm the approximate equivalence, we have also calculated the
expectation values of the plaquette operator $P\!=\! 1 \! - \!
\frac{1}{3}{\rm Re}\Tr U_\Box$ in a variety of different ways.  A pure
gauge calculation at $\beta=5.439$ gives $\Plaq_0 = 0.5190 \pm 0.0004$,
A full QCD hybrid Monte Carlo calculation at $\beta=5.439$ with two
flavors of Wilson fermions gives $\Plaq = 0.4451 \pm 0.0004$.  Our
approximation using the $\beta=5.7$ valence calculation gives $\Plaq_1 =
0.4499 \pm 0.0004$ with a systematic error of $0.0000 \pm 0.0025$.  The
fact that our estimate of the systematic error has a central value of
$0.0$ is simply a consequence of our algorithm for finding $\delta\beta$
since we have chosen $\delta\beta$ explicitly to make this central value
equal to $0.0$.  Other observables will have non-zero systematic shifts
which can be calculated in exactly the same manner.

\section{COMMENT}


The major question which we have not answered yet is whether our scheme
is more efficient computationally than simulating with an exact
full QCD algorithm.  In our simple test the extra work required
to evaluate $\Tr\log\MdM$ turned out to be about equivalent to
the time required to generate a single decorrelated configuration
with the Cabbibo-Marinari algorithm.  How this might change as
the size of the lattice increases, and as the gauge coupling and quark
hopping parameters become more physical is still to be determined.

\section{ACKNOWLEDGEMENTS}

We are grateful to Charles Thorn for several discussions which
stimulated much of the work described here.


\begin{thebibliography}{9}

\bibitem{Sexton94}
J. Sexton, and D. Weingarten, preprint hep-lat/9411029,
November 1994, 24 pages.

\bibitem{Fox68}
L. Fox, and I. B. Parker, Chebyshev Polynomials in Numerical Analysis,
Oxford University Press, London, 1968

\bibitem{Golub89}
G. H. Golub and C. F. Van Loan, Matrix Computations, The Johns Hopkins
University Press, Baltimore, 1989

\end{thebibliography}
\end{document}